\documentstyle[pre,aps,multicol,epsf]{revtex}
\begin{document}
\title{Exchange Effects in the Invar Hardening: $Fe_{0.65}Ni_{0.35}$ as a
test case}
\author{M. Molotskii and V. Fleurov}
\address{
School of Physics and Astronomy, Beverly and Raymond Sackler
Faculty of Exact Sciences.\\ Tel Aviv University, Tel Aviv 69978, Israel}
\date{\today}
\maketitle
\begin{abstract}
An increase of the critical resolved shear stress of Invar alloys
(Invar hardening) with a lowering temperature is explained. The
effect is caused by a growth of the exchange interaction between
dangling $d$-electron states of dislocation cores and paramagnetic
obstacles (e.g., Ni atoms in FeNi alloys) which occurs below the
Curie temperature. The spins of the two electrons align along the
magnetization due to the exchange interaction with the surrounding
atoms of the ferromagnetic. The exchange interaction between the
dislocations and obstacles is enhanced in Invars due to a strong
growth of the magnetic moments of atoms under the action of
elastic strains near the dislocation cores. Parameters
characterizing the exchange interaction are determined for the
case of the Fe$_{0.65}$Ni$_{0.35}$ Invar. The influence of the
internal magnetic field on the dislocation detachment from the
obstacles is taken into account. The obtained temperature
dependence of the critical resolved shear stress in the
Fe$_{0.65}$Ni$_{0.35}$ Invar agrees well with the available
experimental data. Experiments facilitating a further check of the
theoretical model are suggested.
\end{abstract}

\section{Introduction}

It was in 1897 when Guillame first revealed unusual properties of
ferromagnetic FCC iron-nickel alloys with the content close to
Fe$_{0.65}$Ni$_{0.35}$. The alloys appeared to be characterized by
an extremely low thermal expansion in a wide range around the room
temperature. This is because of this property of having an
invariant volume under varying temperature that this alloys have
been named Invars. It was shown later on that many other physical
and mechanical properties of Invars also strongly differ from
those of other ferromagnetics. We mention here among the others:
volume magnetostriction, specific heat, elastic constants,
electrical resistivity, pressure dependence of the magnetization
and the Curie temperature. Such an anomalous behavior is observed
only in the nearest vicinity of the content specific for Invars
and may strongly depend on the temperature, pressure, magnetic
field and so on (see, e.g., review \cite{w90}, and proceedings of
the centennial symposium on the Invar effect \cite{invar97}). Many
unusual properties of Invars can be understood if one considers a
possibility of continuous transitions between the collinear
ferromagnetic spin state to a non-collinear configuration
(\cite{akhsk96,saj99} and references therein).

Plastic properties of Invars are also anomalous. It is known for
long (see, e.g., \cite{h83}) that the critical resolved shear
stress (CRSS) of any alloy grows with lowering temperature. This
growth is explained  by a decreasing probability of dislocation
detachment from obstacles induced by thermal fluctuations. The
same trend is observed in Invars but this growth in Invars is at
least order of magnitude stronger than in non-magnetic FCC alloys,
and several times stronger than even in FeNi alloys with non-Invar
contents \cite{br69,ehy72,eh79,fgs80,rss85,r87,gs96}. It is
emphasized that this anomalously strong temperature dependence of
CRSS in Invar alloys, named Invar hardening, starts only below the
Curie temperature. This fact indicates an active role of the
magnetic order in formation of plastic properties of Invars.

Unique properties of Invars resulted in their extremely broad
application in numerous fields of technology. This certainly
creates a special interest in studying their plastic properties.
Several models suggested for an explanation of the Invar hardening
will be critically discussed in Section II of this paper. We try
to demonstrate in that section that these models are unfortunately
not sufficient for understanding the effect.

Section III describes the model which we propose for an
explanation of the Invar hardening. It is based on specific
features of the exchange interaction between the electrons forming
bonds of dislocations with obstacles. This exchange interaction in
Invars appears to be much more important than widely discussed
elastic and electrostatic interactions.

\section{Models of Invar Hardening}

All the models, proposed with the aim to explain the Invar
hardening, take into account the anomalously strong dependence of
the local magnetization of Invars near edge dislocations on the
local pressure. Krey \cite{k69} showed that the direct interaction
between local magnetizations near the dislocations is weak. It is
three orders of magnitude smaller than the corresponding elastic
interaction. Other magnetic mechanisms, e.g., magnetic interaction
of the dislocations with defects, are also very weak. That is why
all the models disregard magnetic mechanisms and account only for
exchange interactions varying with a change of the Invar
magnetization.

It was proposed by Echigoya et al \cite{ehy72} that Invar
hardening is caused by the so called magnetic friction in the
course of dislocation motion. According to this model the part of
exchange energy, associated with a dislocation, dissipates when
the dislocation moves. It leads to an additional friction, which
does not depend on the dislocation velocity. However, there are
some doubts as for the effectiveness of this mechanism. The
dislocation energy, including its exchange self energy, is
conserved when moving in a translationally invariant crystal. A
dynamical deceleration of dislocations is possible only due to its
interaction with various elementary excitations, such as phonons,
conduction electrons, spin waves and so on. Such processes lead to
a viscous friction which increases with the dislocation velocity
and becomes zero for resting dislocations. There is also a
magnetic friction in ferromagnetics caused by interaction of
dislocations with magnons. The calculation \cite{kp78} shows that
this friction is extremely small even when the dislocation
velocities are close to the speed of sound. As for the slowly
moving dislocations observed in experiments
\cite{br69,ehy72,eh79,fgs80,rss85,r87,gs96} (their velocity is 8
to 9 orders of magnitude smaller than the sound velocity), for
them the magnetic friction is negligibly small and cannot anyhow
affect the plasticity.

Flor et al \cite{fgs80} connect the Invar hardening with the
increasing activation energy for the kink motion, which is caused
by an increase of their exchange energy. It is, however, known
\cite{h83} that the motion of a dislocation in any FCC alloy is
controlled by the dislocation detachment from the obstacles rather
than by the kink motion. This makes the starting point of the
model \cite{fgs80} not quite reliable.

An interesting model for the Invar hardening was proposed by Retat
\cite{r87} who assumed that the spins of the electrons localized
near the dislocation core are antiparallel to the spins of the
electron in the neighboring atoms. This spin orientation realizes,
according to \cite{r87}, the minimum of the energy. If the
dislocation moves to the neighboring equilibrium state so swiftly
that spins do not have time to relax, then the core spins become
parallel to those of the neighboring atoms. The exchange
interaction results in an increase of the dislocation energy and
in an effective growth of the Peierls barrier. This explains,
according to Retat, the Invar hardening.

This mechanism can be effective if dislocation velocities are, at
least, about several cm/s. Such and even larger velocities are
achievable only under strong pulse loads of pure metals when the
dislocation deceleration is due to a viscous friction \cite{ai89}.
However, the experiments
\cite{br69,ehy72,eh79,fgs80,rss85,r87,gs96} are carried out under
static loads and only slowly moving dislocations appear due to
their thermally activated detachment from obstacles. Simple
estimates lead to the conclusion that their velocities are less
than 10$^{-4}$cm/s, i.e., four orders of magnitude smaller than
required for the mechanism \cite{r87} to work.

Domain walls also play a part in pinning dislocations in Invars.
Magnetostriction near the domain walls of the ferromagnetics
results in long range strains, \cite{t69}, hindering the
dislocation motion. That is why creation of domains at
temperatures below the Curie temperature may lead to an increase
of CRSS in ferromagnetics, and might have, in principle, produced
the Invar hardening.

Unfortunately, there are several reservations as for the role of
the domain walls in the Invar hardening. First, the additional
strains produced by the walls ($\sim 1$ MPa) are two orders of
magnitude smaller than the CRSS observed in Invars
\cite{br69,ehy72,eh79,fgs80,rss85,r87,gs96}. Second, these strains
are proportional to the coefficients of the linear
magnetostriction, and in Invars these coefficients do not
essentially differ from those in other ferromagnetics \cite{lb62},
meaning that the fact that this type of hardening is specific only
for Invars cannot be understood in this way. Third, the Invar
alloys are soft magnetic materials with low values of the coercive
field \cite{cw80}. This indicates a weak pinning of the domain
walls by all the lattice defects including the dislocations.
Fourth, direct measurements \cite{r87} of the influence of the
domain walls on CRSS in Invars do not favor their participation in
the Invar hardening.

Domains characterized by the magnetic momentum along the external
magnetic field grow. This grows diminishes the number of domains
walls pinned by dislocations. Therefore, the dislocation path
length also grows in the magnetic field. This leads to an
increasing plasticity of the ferromagnetic in a rather weak
($\sim$ 0.01 T) magnetic field \cite{h72}. However, the
experiments \cite{r87} show that the 0.04 T field, quite
sufficient to essentially reduce the number of domain walls, does
not have any influence on CRSS. We have no other choice as to
conclude that domain walls hardly play any part in the Invar
hardening.

We see that existing models do no provide a convincing explanation
of the plastic properties of Invars. We will try in the next
section to formulate an alternative model. The important feature
of this model is a strong temperature dependence of the exchange
interaction between dislocations and solute atoms (Ni atoms in
case of FeNi alloy) which allows one to explain the anomalous
plastic properties of Invars.

\section{
Role of the Dislocation -- Obstacle Exchange Interaction in the
Invar Hardening}

We are going to present in this section our explanation of the
effect of Invar hardening. When carrying out the discussion and
numerical estimates we will mainly refer to the
Fe$_{0.65}$Ni$_{0.35}$ alloy, although our model with minor,
mainly numerical, corrections is applicable to Invars with other
contents as well.

The theory of plastic properties of alloys usually considers two
main types of the dislocation interaction with defects --- elastic
and electrostatic \cite{h83}. The elastic interaction arises if the
size of the impurity atoms differs strongly from the host atoms
and a strain field is created in their vicinity. Then they
interacts with the strain fields of dislocations. The
electrostatic interaction in metals is due to the electric fields
created due to a redistribution of charges near a dislocation,
which acts on the charge distributions near defects.

Both types of the interaction are expected to be very weak in
Invars, since atoms constituting them have very close sizes and
chemical properties. For example in Fe$_{0.65}$Ni$_{0.35}$ Invar,
the size of Ni (1.24\AA) is hardly distinguishable from that of Fe
(1.28\AA\ in FCC lattice). Therefore, one should not expect any
strain field around Ni atoms and, no tangible elastic interaction
of dislocations with Ni atoms is possible. Ni atoms do not also
create additional charges, since they are characterized by the
same valence and electro-negativity as Fe atoms, making the
electrostatic interaction between the dislocations and Ni atoms
also negligible.

However, Invar alloys are characterized by a strong spin
inhomogeneity. Calculations \cite{p89,ehmsm93,seaehr95} of the
density of states with oppositely directed spins demonstrate that
different atoms in Invar alloys possess unsaturated d-states with
differing values of the localized magnetic moments. For example,
Fe atoms in Fe$_{0.65}$Ni$_{0.35}$ Invars have a 2.5$\mu_B$
magnetic moment, in contrast to 0.7$\mu_B$ typical of Ni atoms
\cite{yis79}; here $\mu_B$ is the Bohr magneton. As a result,
localized spin states are associated with Ni atoms. The authors
\cite{aggkt96} consider, so called, "chemical" interaction of the
dangling $d$-bonds of a dislocation cores and impurities. Their
calculation shows that the interaction can be very strong, up to 1
eV. This type of interaction is specific for the edge dislocations
and is absent in the case of the screw dislocations. The screw
dislocations can hardly form such bonds, which manifests itself in
their much higher mobility observed in the experiments
\cite{eh79,rss85}. The structure of chemical bonds is known to be
largely controlled by the exchange interaction (see, e.g.,
\cite{hw95}). It is the exchange energy which makes the difference
between the chemical bonds formed by electrons with parallel and
antiparallel spins. Depending on the sign of the exchange
interaction - positive (ferromagnetic) or negative
(antiferromagnetic) - the bond with either parallel or
antiparallel spins becomes bonding, whereas the chemical bond with
the opposite spin configuration becomes antibonding.

We have recently demonstrated
\cite{mkf95,mf95,mf96,mf97a,mf97b,mf00,m00} that considering the role of
the exchange interaction in formation of the bonds between obstacles
and edge dislocations plays a crucial role in the dislocation
dynamics under the influence of a magnetic field. Such an approach
allowed us, in particular, to understand the microscopic
mechanisms of the electro- and magnetoplastic effects. Here we are
going to apply similar ideas for an explanation of the Invar
hardening.

The exchange interaction between the electrons in the dangling bonds of
the dislocation core and the solute atom can be conventionally described as
\begin{equation}
U_{dk} = - 2 J ({\bf S}_{d} {\bf S}_{k}) \label{exchange1}
\end{equation}
where $J$ is the exchange integral, while ${\bf S}_{d}$ and ${\bf
S}_{k}$ are spin vectors related to the dislocation and defect
electron spins. In a magnetic medium these vectors tend to orient
along the common magnetization vector. Then using the molecular
field approximation one can substitute them by their average
values and rewrite equation (\ref{exchange1}) as
\begin{equation}
U_{dk} = - U_{0} - 2 J S_{d} S_{k} \sigma^{2}(T).
\label{exchange2}
\end{equation}
where $\sigma (T)$ is the temperature dependent relative
magnetization.

The term, $U_0$, of the interaction (\ref{exchange2}) does not
disappear even at $T>T_{c}$. It is a smooth function of the
temperature near $T_{c}$ and accounts for the correlations between
the orientations of the interacting spins which remain nonzero
even in the paramagnetic state. It is quite sufficient for our
purposes to represent this smooth function by a constant $U_{0}$.
A similar equation has been deduced in reference \cite{m85}.

Interaction between defects can be also described by this type of
equation. For example, the interaction between vacancies in
$\alpha $-Fe grows from 0.24 to 0.30 eV when transiting from the
para- to the ferromagnetic state \cite{d75}. This growth is well
described when assuming interaction potential (\ref{exchange2})
with $U_{0}=0.24$ eV and $2JS_{d}S_{k}=0.06$ eV.

It is specific for the Invar alloys that the magnetic moments of
atoms depend anomalously strongly on the pressure
\cite{ks60,hm81}, decreasing under a positive pressure and
increasing under a negative pressure. The dangling bonds of the
edge dislocations are situated in the region of the missed
semi-plane, i.e., they are under the action of a negative
pressure. The same can be said about the $Ni$ atoms whose
electrons interact with the dislocation dangling bonds. The
negative pressure results in a growth of both values of the spins
$S_d$ and $S_k$ and, hence, the linear approximation for the
pressure dependence of the atomic moments near the dislocation
axis, results in
\begin{equation}
S_{i}=S_{i}^{0}(1+\alpha p) \label{moment}
\end{equation}
with $i=d,k$. Here $S_{i}^{0}$ is the spin value at zero pressure
$p$. $\alpha $ is a material constant. In the region of the missed
semi-plane the pressure near the dislocation is \cite{f64}
\begin{equation}
p = -\frac{\mu (1+\nu )} {3\pi (1-\nu )}  \label{pressure}
\end{equation}
where $\mu$ is the shear modulus and $\nu$ is the Poisson
coefficient. Now substituting equations (\ref{moment}) and
(\ref{pressure}) into (\ref {exchange2}) one gets the exchange
interaction of dislocations with atoms in the form
\begin{equation}
U_{dk} = -U_{0}-\delta U g^{2} \sigma^{2}(T)  \label{exchange3}
\end{equation}
where
\[
\delta U = 2 J S_{d}^{0} S_{k}^{0},
\]
and
\[
g=1-\frac{\alpha \mu (1+\nu )}{3\pi (1-\nu )}.
\]

When considering conventional ferromagnetics one can describe the
relative magnetization using the Curie-Weiss model according which
\begin{equation}\label{CW}
\sigma (T)=\left\{
\begin{array}{cc}
\sqrt{1-\displaystyle\frac{T}{T_{c}}}, & \mbox{at }T\leq T_{c} \\
0, & \mbox{at }T>T_{c}
\end{array}
\right.
\end{equation}
with $T_{c}$ being the Curie temperature. However, the temperature
dependence of the relative magnetization $\sigma(T)$ in Invars
deviates from the Curie-Weiss equation (\ref{CW}). This deviation
is caused by the coexistence of several magnetic states whose
energies are very close to the energy of the ground ferromagnetic
state \cite{saj99}. The most important deviation occurs near the
Curie temperature where $\sigma(T)$ does not go to zero but
acquires small (0.15 to 0.20) values and rapidly decreases at
higher temperatures \cite{w90}. Here we are interested in the
range below the Curie temperature where we expect an influence of
the magnetic ordering on plasticity. In this temperature range
using equation (\ref{CW}) is a rather rough but reasonable
approximation. Even at higher temperatures close to $T_C$ this
approximation does not cause an essential error since the relative
magnetization is small and the energy (\ref{exchange2}) is
proportional to $\sigma^2(T)$, making the corresponding correction
even smaller.

According to (\ref{exchange3}) the temperature dependence of the
interaction energy is the stronger the more the value of $g$ deviates
from one. This value essentially differs from one only for Invar alloys with
anomalously large parameter $\alpha $. For example, the
Fe$_{0.65}$Ni$_{0.35}$ alloy is characterized by $\alpha =-1.1
\times 10^{-11}$ (dyn/cm$^{2})^{-1}$ \cite{hm81}. Using the values
$\mu =6.35 \times 10^{11}$ dyn/cm$^{2}$ and $\nu =0.3$, one gets
$g=2.38$. It is known \cite{ks60}\ that the slightest deviation of
the alloy content from the composition Fe$_{0.65}$Ni$_{0.35}$
results in a decrease of the absolute value of the parameter
$\alpha$ by an order of magnitude or even more. As a result
$g\approx 1$ in non-Invar compositions and, hence, for these
alloys the potential (\ref {exchange3}) depends on the temperature
much weaker. We believe that this is the principle cause of the
anomalously strong temperature dependence of CRSS in Invar alloys.

According to the data available in reference \cite{fgs80} the
value $\tau _{c}$ of CRSS depends on the Ni concentration as
$c_{at}^{2/3}$. Such a dependence corresponds to the solid
solution hardening model proposed by Labush \cite{l70} according
which the CRSS at $T=0$ (i.e., the stress necessary to move a dislocation
through a random array of obstacles in the glide plane) is
\begin{equation}
\tau _{c0} = \frac{c_{at}^{2/3} f_{m}^{4/3} w^{1/3}C} {a^{4/3}
b(4T_{l})^{1/3}}. \label{crss1}
\end{equation}
Here $b$ is the magnitude of the Burgers vector, $T_{l}\approx
\frac{1}{2}\mu b^{2}$ is the line tension of the dislocation,
$a^{2}$ is the area of a lattice cell, $w$ is the range of the
dislocation - obstacle interaction force, $f_{m}=U_{dk}/w$ is the
maximal interaction force; $C=0.36000$ for a repulsive
interaction, and $C=0.27956$ for an attractive interaction.

If one also accounts for thermal activation processes, equation
(\ref{crss1}) becomes \cite{h83}
\begin{equation}
\tau_{c}(T) = \frac{c_{at}^{2/3} f_{m}^{4/3} w^{1/3}C} {a^{4/3} b
(4T_{l})^{1/3}} \left[ 1-\left( \frac{T}{T_{0}}\right) ^{
{\displaystyle{2 \over 3}} }\right] ^{ {\displaystyle{3 \over 2}}
},  \label{crss2}
\end{equation}
where $T_{0}$ is a characteristic temperature at which
$\tau_{c}(T)$ becomes zero. The characteristic temperature $T_{0}$
is proportional to the interaction energy between the dislocation
and the obstacle \cite{h83}. As shown above (\ref{exchange3}),
this energy itself depends on the temperature in the ferromagnetic
state and, hence, the characteristic "temperature" may be
represented as
\[
T_{0}(T)=\left\{
\begin{array}{lc}
T_{0p}\left[ 1+ {\displaystyle{\delta U \over U_{0}}}
g^{2}\sigma^{2}(T)\right], & \mbox{at }T\leq T_{c} \\ T_{0p}, &
\mbox{at }T>T_{c}
\end{array}
\right.
\]
where $T_{0p}$ is the value of the parameter $T_{0}$ in the
paramagnetic state.

Now we are in a position to consider the influence of the internal
magnetic field on plastic properties of Invar alloys. There is a
spontaneous magnetization M of about 1 to 2$\times 10^6$A/m which
corresponds to an internal magnetic induction $B\sim 1\ {\rm to}\
2$T. This field is strong enough to influence plastic properties
of crystals (see our papers \cite{mkf95,mf95,mf96,mf97a,mf97b,mf00,m00}
and references therein). It has been demonstrated in these papers
that effective concentration of obstacles participating in the
dislocation pinning depends on the magnetic field. We believe that
a similar mechanism is responsible for the influence of varying
magnetization on the CRSS in Invars.

According to this approach the binding energy of the bond between
a dislocation and a paramagnetic obstacle strongly depends on its
multiplicity. In the paper \cite{mf97b} we assumed that the singlet
state of the bond possesses energy which is lower than in a triplet state,
which corresponds to the negative (antiferromagnetic) sign of the
parameter $J$ in equation (\ref{exchange1}). According to our model
an external magnetic field may induce singlet to triplet transitions
and lead to an increased population of the $T$ state with lower binding
energy and, hence, to an increased  probability of the dislocation
detachment from the obstacles. The crystal plasticity may, as a
result, grow.

The Invar alloys which we discuss now are ferromagnetics and it is
more natural to assume that the exchange parameter $J$ in equation
(\ref{exchange1}) is now positive (ferromagnetic). Then the
triplet state will have a larger binding energy, and magnetic
field induced transitions to the singlet state will facilitate
detachment of the dislocations. This process can be described by
the same equations as in \cite{mf97b}. The opposite sign of the
exchange energy does not have any consequences for these
equations, since they relate to the dynamics of the system in the
resonance region, where the exchange energy is negligible. But,
contrary to the antiferromagnetic case, we assume that the
dislocation bond is initially formed in one of three $T$ states
with equal probability (1/3). We should also keep in mind that the
difference between the magnetic field $\mu_0 H$ and the induction
$B$ is important in ferromagnetics ($\mu_0=4\pi\times
10^{-7}$kg$\cdot$m$^3$sec$^{-4}$A$^{-2}$ is magnetic constant).
Solving equations presented in reference \cite{mf97b}, we look for
the probability $\rho_T(B)$ that, on having passed the resonance
region under the action of a magnetic induction $B$, the system
remains in one of the three triplet states:
\begin{equation}
\rho_T(B)=\frac{1}{1+\displaystyle{\frac{B^2}{B_0^2}}}
\label{triplet}
\end{equation}
where $$B_0=\sqrt{6\displaystyle{(1 +
\frac{T_1}{\tau})(1+\frac{T_2}{\tau})}}B_m,$$
$$B_m=\displaystyle{\frac{\hbar}{\Delta g\mu_B\sqrt{T_1T_2}}}.$$
Here $\Delta g$ is the difference of the $g$ factors of the two
electrons forming the radical pair (dislocation -- obstacle bond);
$T_1$ and $T_2$ are the transversal and longitudinal relaxation
times characterizing the radical pair. The typical value of the
characteristic field $B_0$ can be estimated to be from several
tenths to one Tesla.

An obstacle is capable of pinning a dislocation in a ferromagnetic only if
the bond between them is in a binding triplet state. If the radical pair is
excited to its singlet state, the dislocation passes such an obstacle freely
without "seeing" it. One may say that the effective concentration $c_{ac}(B)$
of active obstacles is proportional to $\rho_T(B)$. This concentration,
$c_{ac}(B)$, is now to be used instead of $c_{at}(B)$ in equation
(\ref{crss2}) The temperature dependence of this effective concentration
arrises due to the temperature dependence of the local magnetic field which
in the Lorentz model reads
$B_M(T)=\displaystyle{\frac{1}{3}\mu_0 M(0)\sigma(T)}$. Substituting it into
equation (\ref{triplet}) one gets
\begin{equation}
c_{ac}(T)=\frac{c_{ac}(0)}{1+\left(\displaystyle{\frac{\mu_0
M(0)\sigma(T)}{3B_0}}\right)^2} \label{concentr}
\end{equation}
where $\mu_0 M(0)$ is the saturation magnetization of the
ferromagnetic at $T=0$.

Comparing equation (\ref{concentr}) with our results obtained
earlier \cite{mkf95,mf95,mf96,mf97a,mf97b,mf00,m00} one can conclude
that the effective concentration of the active obstacles does not
in fact depend on the sign of the exchange parameter $J$.
Therefore, the fact that the Invar magnetic ordering is not purely
ferromagnetic but rather a superposition of collinear and
noncollinear states \cite{akhsk96,saj99} does not play a role
here. This rather complex structure produces a temperature
dependent local magnetic moment which is to be used in equation
(\ref{concentr}).

Equations (\ref{crss2}) and (\ref{concentr}) determine the
temperature dependence of the CRSS. If we take the
$Fe_{0.65}Ni_{0.35}$ Invar alloy, then $T_c=503$K \cite{w90},
$\mu_0 M(0)=0.14$T \cite{k69}, $b=2.5\times10^{-10}$m,
$w=\sqrt{3}b=4.3\times 10^{-10}$m, $a^2=\sqrt{3}b^2/3=3.6\times
10^{-20}$m$^2$, $C=0.27956$. The values of the other parameters:
$T_{0p}=980$K, $U_0=0.15$eV, $\delta U=0.10$eV, and $B_0=0.36$T,
are obtained by fitting equation (\ref{crss2}) to the experimental
data \cite{fgs80}.
\begin{figure}[htb]
\epsfysize=19\baselineskip \centerline{\hbox{
\epsffile{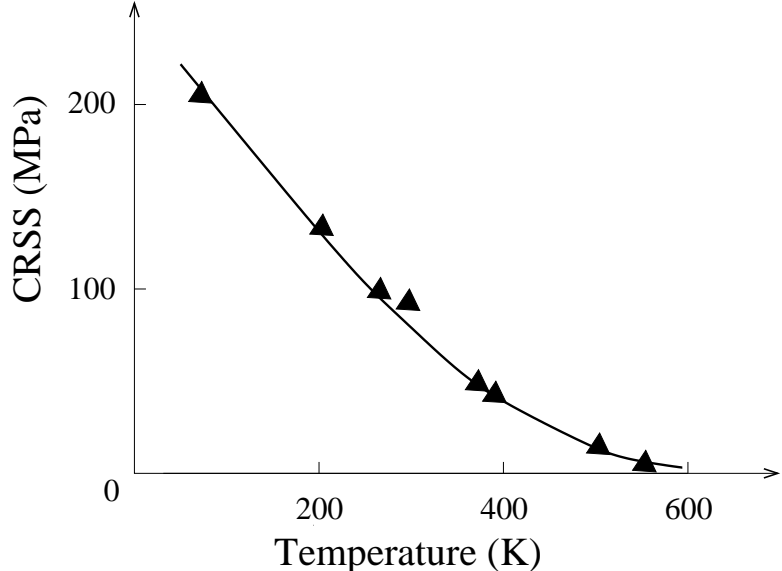}}} \caption{Comparison of the theory and the
experiment for the temperature dependence of the critical resolved
shear stress (CRSS) for Fe$_{0.65}$Ni$_{0.35}$ Invar alloy. The solid
line is calculated my means of equation (\ref{crss2}). The
experimental data are taken from the paper by Flor et al
{\protect\cite{fgs80}}.}
\end{figure}
Fig. 1 exhibits a good agreement of the theory and experiment for
these values of the parameters. It is worth mentioning that the
values itself are also quite reasonable. The above characteristic field
$B_0=0.36$T, which corresponds to the typical values of the magnetic
induction at which dislocations effectively detach from obstacles, lie
 between the value $B_0=0.49$T obtained for aluminum \cite{mf95,mf96} and
$B_0=0.22$T for copper \cite{mkf95}, The $U_0$ and $\delta U_0$ values
are close to those (0.24 and 0.06 eV) obtained for vacancies in $\alpha$-Fe
\cite{d75}. The absolute value of $U_{dk}$ varies at low
temperature within the range from 0.15 to 0.72 eV, which coincides
with the typical range of the energies of the dislocation --
obstacle interaction \cite{f64}.

\section{Conclusions}

As has been demonstrated in the above section the Invar hardening
can be explained by the temperature dependence of the exchange
interaction energy in the system of dislocations bound to
obstacles. Estimates for Fe$_{0.65}$Ni$_{0.35}$ Invars show that
at reasonable values of the parameters there is a quantitative
agreement between the experimental data and the theory.

It has been emphasized in the previous section that the main
characteristics which really strongly distinguish between Fe and
Ni atoms are their magnetic moments. This has lead us to the
conclusion that a special attention should be paid to the exchange
energy contribution to the edge dislocation -- obstacle
interaction. It is demonstrated in this paper that our approach
provides a reasonable quantitative description of the Invar
hardening.

Our model represents, as far as we know, the first attempt to
consider the influence of the internal magnetic field on
plasticity. Since the major part of construction materials are
various steels, which are ferromagnetics, these results may be of
an importance for developing a microscopic theory of plastic
properties of steels.

There are also several other consequences of this model which,
being experimentally verified, may serve as additional independent
confirmation of the validity of our model. These are:
\begin{enumerate}
\item
CRSS grows with a decreasing temperature due to the increase of
the dislocation -- obstacle interaction $U_{dk}$. We expect that
such a behaviour of $U_{dk}$ should also show up in the amplitude
- independent internal friction of dislocations. It was shown in
reference \cite{f84} that the internal friction $Q^{-1}$ in alloys
varies inversely proportional to $U_{dk}$. That is why, one may
expect a rapid fall of the quantity $Q^{-1}$ at $T<T_c$, when the
energy $U_{dk}$ starts growing (see fig. 2). The temperature
dependence of the internal friction $Q^{-1}$ is expected to be
stronger than linear in contrast to conventional nonmagnetic
alloys.
\item
This paper connects the Invar hardening with a strong increase of
the magnetic moments created by the $d$-electrons in atoms in the
vicinity of the dislocation cores. This, as shown in reference
\cite{m84}, causes a strong $s$-$d$-scattering of the conduction
electrons and makes the principal contribution to the dislocation
electric resistivity. Hence, one may expect that the anomalous
increase of the magnetic moments of atoms near the dislocation cores
will result in an anomalously strong dislocation electric
resistivity, as compared to other ferromagnetic metals.
\item
A strong temperature dependence of CRSS in the
Fe$_{0.65}$Ni$_{0.35}$ Invar alloy is explained by a large value
of the parameter $\alpha$ in the dependence of the magnetic
moments on the pressure. Another Invar alloy --
Fe$_{0.72}$Pt$_{0.28}$ -- is characterized at the room temperature
by the parameter $\alpha$ which is factor of 2.2 larger than in
Fe$_{0.65}$Ni$_{0.35}$. Hence, the Invar hardening in this alloy
may be expected to by much stronger than in
Fe$_{0.65}$Ni$_{0.35}$.
\item
The model proposed in this paper demonstrates the importance of
the internal magnetic field in formation of the plastic properties
of ferromagnetics. Strong fluctuations of the local magnetic field
due to the fluctuations of the magnetization near the Curie
temperature (see, e.g., \cite{m76}) may cause an enhanced
detachment of dislocations from obstacles and an increase of
plasticity at this temperature. The fluctuations of the
magnetization on the paramagnetic side of the transition are twice
as large as on the ferromagnetic side \cite{m76}.
\begin{figure}[htb]
\epsfysize=19 \baselineskip \centerline{\hbox{
\epsffile{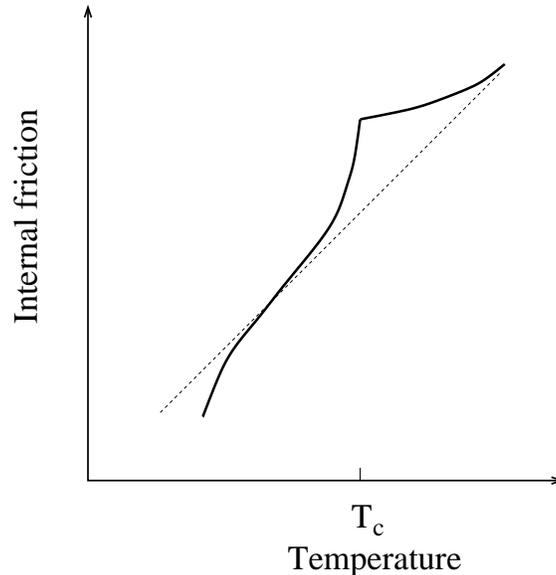}}} \caption{ The expected qualitative shape
of the temperature dependence of the amplitude independent
internal friction of dislocations in Invars --- solid line. The
dashed line shows this dependence in nonmagnetic alloys.}
\end{figure}
In Invars, characterized by an anomalously large magnetic
susceptibility, these fluctuations may be especially strong on both
sides of the phase transition temperature.
Hence, asymmetric peaks, with a larger high temperature shoulder,
may appear in the temperature dependence of the plasticity of
Invars near the Curie temperature. The internal friction of
dislocations, which strongly depends on the magnetic field
\cite{mkf95}, may serve as a convenient technique of studying such
peaks. One may expect asymmetric peaks in the temperature
dependence of the internal friction near the Curie temperature
(see fig.2).
\end{enumerate}

\end{document}